\documentclass[aps,prl,unsortedaddress,twocolumn,superscriptaddress]{revtex4-1}

\usepackage{amsmath}
\usepackage{hyperref}
\usepackage{graphicx}
\usepackage{color}

\begin{document}
\title{Formation of buried domain walls in the ultrafast transition of SmTe$_3$}

\author{M. Trigo}
\email[E-mail: ]{mtrigo@slac.stanford.edu}
\affiliation{Stanford PULSE Institute, SLAC National Accelerator Laboratory, Menlo Park, CA 94025, USA}
\affiliation{Stanford Institute for Materials and Energy Sciences, SLAC National Accelerator Laboratory, Menlo Park, CA 94025, USA}

\author{P. Giraldo-Gallo}
\affiliation{Department of Applied Physics, Stanford University, Stanford, CA 94305, USA}
\affiliation{Department of Physics, Universidad de Los Andes, Bogot\'a, 111711, Colombia.}

\author{J. N. Clark}
\affiliation{Stanford PULSE Institute, SLAC National Accelerator Laboratory, Menlo Park, CA 94025, USA}

\author{M. E. Kozina}
\affiliation{Stanford PULSE Institute, SLAC National Accelerator Laboratory, Menlo Park, CA 94025, USA}
\affiliation{Stanford Institute for Materials and Energy Sciences, SLAC National Accelerator Laboratory, Menlo Park, CA 94025, USA}
\affiliation{Department of Applied Physics, Stanford University, Stanford, CA 94305, USA}

\author{T. Henighan}
\author{M. P. Jiang}
\affiliation{Stanford PULSE Institute, SLAC National Accelerator Laboratory, Menlo Park, CA 94025, USA}
\affiliation{Stanford Institute for Materials and Energy Sciences, SLAC National Accelerator Laboratory, Menlo Park, CA 94025, USA}
\affiliation{Department of Physics, Stanford University, Stanford, CA 94305, USA}

\author{M. Chollet}
\affiliation{Linac Coherent Light Source, SLAC National Accelerator Laboratory, Menlo Park, California 94025, USA}

\author{I. R. Fisher}
\affiliation{Stanford Institute for Materials and Energy Sciences, SLAC National Accelerator Laboratory, Menlo Park, CA 94025, USA}
\affiliation{Department of Applied Physics, Stanford University, Stanford, CA 94305, USA}

\author{J. M. Glownia}
\affiliation{Linac Coherent Light Source, SLAC National Accelerator Laboratory, Menlo Park, California 94025, USA}
\author{T. Katayama}
\affiliation{Japan Synchrotron Radiation Research Institute, 1-1-1 Kouto, Sayo-cho, Sayo-gun, Hyogo 679-5198, Japan}
\author{P. S. Kirchmann}
\affiliation{Stanford Institute for Materials and Energy Sciences, SLAC National Accelerator Laboratory, Menlo Park, CA 94025, USA}
\author{D. Leuenberger}
\affiliation{Stanford Institute for Materials and Energy Sciences, SLAC National Accelerator Laboratory, Menlo Park, CA 94025, USA}
\affiliation{Department of Applied Physics, Stanford University, Stanford, CA 94305, USA}
\author{H. Liu}   
\affiliation{Stanford PULSE Institute, SLAC National Accelerator Laboratory, Menlo Park, CA 94025, USA}
\affiliation{Department of Physics, Stanford University, Stanford, CA 94305, USA}

\author{D. A. Reis}
\affiliation{Stanford PULSE Institute, SLAC National Accelerator Laboratory, Menlo Park, CA 94025, USA}
\affiliation{Stanford Institute for Materials and Energy Sciences, SLAC National Accelerator Laboratory, Menlo Park, CA 94025, USA}
\affiliation{Department of Applied Physics, Stanford University, Stanford, CA 94305, USA}

\author{Z. X. Shen}
\affiliation{Stanford Institute for Materials and Energy Sciences, SLAC National Accelerator Laboratory, Menlo Park, CA 94025, USA}
\affiliation{Department of Applied Physics, Stanford University, Stanford, CA 94305, USA}

\author{D. Zhu}
\affiliation{Linac Coherent Light Source, SLAC National Accelerator Laboratory, Menlo Park, California 94025, USA}


\begin{abstract}
We study ultrafast x-ray diffraction on the charge density wave (CDW) of SmTe$_3$ using an x-ray free electron laser. The CDW peaks show that photoexcitation with near-infrared pump centered at 800 nm generates domain walls of the order parameter propagating perpendicular to the sample surface. These domain walls break the CDW long range order and suppress the diffraction intensity of the CDW for times much longer than the $\sim 1$~ps recovery of the local electronic gap. We reconstruct the spatial and temporal dependence of the order parameter using a simple Ginzburg-Landau model and find good agreement between the experimental and model fluence dependences. Based on the model we find that at long times, depending on the pump fluence, multiple domain walls remain at distances of few nm from the surface.

\end{abstract}


\maketitle

A fast quench through a critical point produces topological defects separating domains with distinct values of the order parameter~\cite{kibble1976,zurek1985,zurek1996}. At much faster timescales, topological defects can be created in condensed matter systems with a spontaneously broken symmetry by ultrafast laser pulses~\cite{yusupov2010,zong2019,nasu2004}. Fine control over these defects could provide a route to reach thermodynamically inaccessible~\cite{stojchevska2014,ichikawa2011,kogar2019} or topologically inequivalent  states~\cite{sie2019}, enabling novel forms of control of quantum phases~\cite{basov2017}. 
But imaging the defects as they are produced by ultrafast pulses is a daunting challenge. 
We report ultrafast diffraction experiments using an x-ray free electron laser (XFEL) on SmTe$_3$, a prototypical charge density wave (CDW) material. 
By virtue of the high momentum and time resolution afforded by the XFEL we observe a fast broadening of the diffraction peaks at $0.4$~ps that reflects the creation and coherent dynamics of domain walls of the CDW lattice distortion propagating perpendicular to the sample surface. Aided by a simple model, we reconstruct the depth- and time-dependence of the order parameter and we observe the creation of one, two, or three domain walls depending on the pump excitation strength. 
The potential to produce and visualize defects on demand will advance our understanding of their role in stabilizing other intertwined orders in CDWs~\cite{kogar2019,zhou2019nonequilibrium,hamlin2009} and other quantum materials~\cite{fradkin2015theory}.

Universality and the theory of critical phenomena attest at the success of our understanding of equilibrium second-order phase transitions~\cite{goldenfeld1992lectures}. Much less understood are the dynamics of symmetry-breaking phase transitions away from equilibrium. The Kibble-Zurek mechanism provides a statistical description of the formation of topological defects upon cooling through the critical temperature $T_c$~\cite{kibble1976,zurek1996}. Rather than statistical averages, direct imaging and control of symmetry-breaking defects can potentially enable defects on demand applications~\cite{basov2017} and may enable theoretical developments by observing the coherent dynamics governed by intrinsically non-linear equations of motion~\cite{mcmillan1975,kusar2011,huber2014,trigo2019}. Ultrafast lasers provide an attractive way to create topological defects through fast non-adiabatic excitation and XFELs enable capturing their ultrafast dynamics~\cite{moore2016,trigo2019}. 

\begin{figure*}[htb]
\centering 
\includegraphics[width=6in]{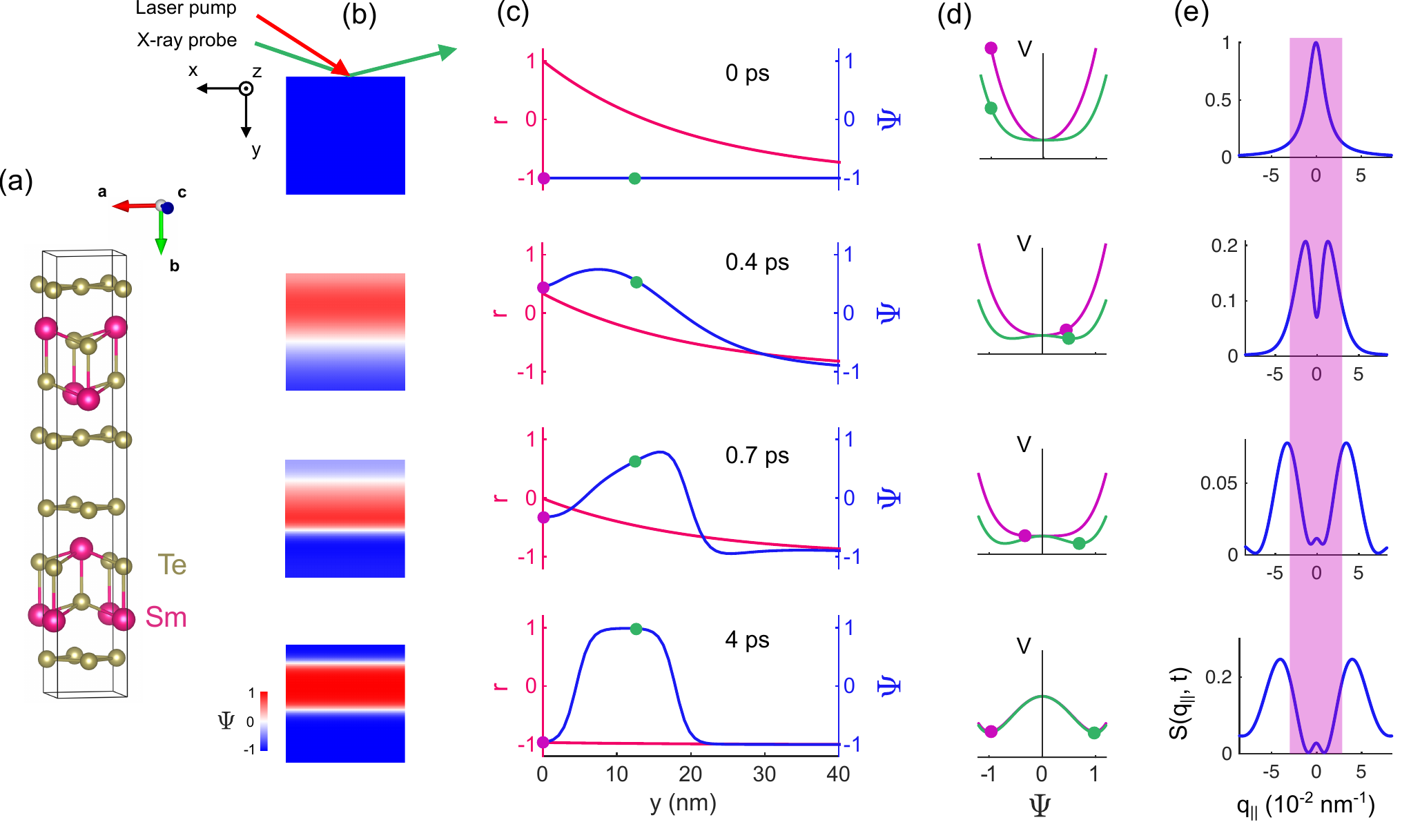} 
\caption[]{(a) The crystal structure of SmTe$_3$ without the CDW distortion. (b) geometry of the experiment and representative snapshots of the dynamics of $\Psi(y,t)$. (c) lineouts of the depth dependence of $\Psi(y,t)$ (blue) and the quadratic coefficient $r(y,t)$ (red). (d) Solid lines show $V(\Psi(y,t))$ and the corresponding value of $\Psi(y,t)$ at delays shown in (c) for $y=0$ (purple dot) and $y=13$~nm (green dot), also indicated in (c). (e) Structure factor $S(q_\parallel, t)$ produced by the corresponding $\Psi(y,t)$ shown on (c). $q_\parallel$ is the wavevector parallel to $y$ with $q_\parallel=0$ the nominal CDW Bragg condition. The shaded area indicates the range of wavevectors probed in the experiment.
\label{fig:1} } 
\end{figure*}

Here we use high-resolution x-ray diffraction at the Linac Coherent Light Source (LCLS) XFEL to resolve fine momentum dependent features in the dynamics of the CDW Bragg peak of SmTe$_3$ from which we infer the coherent evolution of the spatially-dependent order parameter with the help of a simple model. We find that inhomogeneous photoexcitation due to the finite penetration depth of the femtosecond, near-IR pump flips the CDW amplitude resulting in inequivalent regions separated by domain walls perpendicular to the sample normal. The fine time resolution allows us to observe the production and stabilization of one, two, and three domain walls, depending on the excitation fluence.
These domains are long-lived and their diffraction interferes destructively, suppressing the CDW peaks up to nanoseconds~\cite{zhou2019nonequilibrium}, long after the CDW gap has recovered fully~\cite{zong2019}. 

We focus on the CDW in SmTe$_3$ with ordering wavevector $\mathbf{q}_{\mathrm{cdw}} = (0, 0, q)$~rlu (reciprocal lattice units) with $q \approx 2/7$ (see structure in Fig.~\ref{fig:1}a), which develops at $T < T_c = 416$~K. This class of layered materials has recently attracted attention as a model system to investigate the dynamics of symmetry-breaking phase transitions~\cite{yusupov2010,zong2019,kogar2019,zhou2019nonequilibrium,moore2016,trigo2019}.
The disparity between the fast relaxation of the photoexcited charges, of order $\sim 1$~ps~\cite{zong2019,rettig2014,schmitt2011,leuenberger2015}, and the slow recovery of the long-range order in the lattice~\cite{moore2016,trigo2019,zong2019} of order nanoseconds~\cite{zhou2019nonequilibrium}, suggest that the lattice remains non-ergodic long after the electrons have cooled. This has been attributed to the creation of topological defects corresponding to in-plane dislocations of the Te-Te layers~\cite{zong2019}. The near-IR pump also produces significant inhomogeneity of the CDW perpendicular to the layers, as observed indirectly by ultrafast reflectivity~\cite{yusupov2010}. 
Clearly, better visualization of how these defects are created and decay will clarify their topological stability and will yield new insight into how they stabilize other degrees of freedom~\cite{kogar2019,hamlin2009,fradkin2015theory}.

To illustrate the creation of domain walls and their signatures in the diffraction intensity, we consider a minimal one-dimensional model with a real-valued order parameter $\Psi(y, t)$, which represents the CDW lattice distortion in SmTe$_3$. While phase fluctuations are expected in this incommensurate CDW, they take time to develop and do not affect the initial dynamics.
Here, $y$ is the direction perpendicular to the sample surface, as shown schematically in Fig.~\ref{fig:1}a-b. A $\pi$ phase shift in $\Psi$ represents a reversal of the amplitude of the CDW distortion propagating along the $y$ axis (Fig.~\ref{fig:1}a-b). We consider a spatially- and temporally-dependent Ginzburg-Landau potential~\cite{yusupov2010}
\begin{equation}\label{eq:potential}
	V(\Psi) = r(y, t) |\Psi|^2 + \frac{1}{2}|\Psi|^4 + \xi^2 |\nabla \Psi|^2,
\end{equation}
where the third term accounts for the strain energy of a spatially-inhomogeneous configuration~\cite{mcmillan1975}. Here $r = -1$ and $\Psi = \pm 1$ ($r>0$ and $\Psi = 0$) correspond to the CDW ordered (disordered) phase and $\xi = 1.2$~nm is the coherence length~\cite{yusupov2010}.
The coefficient $r(y,t > 0) = \eta e^{-t/\tau} e^{-y/y_p} -1$ represents the sudden photoexcitation on the potential energy, with $\eta$ proportional to the pump fluence~\cite{trigo2019,yusupov2010,huber2014}. Importantly, $r(y,t)$ is spatially inhomogeneous due to the finite penetration depth of the pump, $y_p = 20$~nm.
As we will show next, for sufficiently high excitation, $r > 0$ near the surface, and $\Psi$ can transiently be reversed producing alternating regions with $\Psi = \pm 1$ (Fig.~\ref{fig:1}b). 
When the electronic excitation recovers quickly, i.e. when $\tau$ is fast compared with the dynamics, the inhomogeneities in $\Psi$ are frozen leaving behind domain walls.

We integrate the equation of motion derived from the potential in Eq.~(\ref{eq:potential}) numerically (see Supplementary Information). Fig.~\ref{fig:1}c shows $\Psi(y,t)$  (blue curve, right axis) and $r(y,t)$ (red curve, left axis) at representative times for $\eta = 2$ (see Supplementary Movies). Fig.~\ref{fig:1}d shows the potential $V(y,t)$ for two representative depths, $y=0$ (purple curve) and $y=13$~nm (green curve). 
Initially, $r(y=0,t=0) = 1$ and the potential is strongly harmonic at $y=0$ (purple line in (d)) and $\Psi(y=0, t)$ acquires significant potential energy, (purple dot). On the other hand, at $y=13$~nm, $r \approx 0$, the potential is mostly quartic (green curve) and $\Psi$ has less potential energy (green dot). 
At $t=0.4$~ps the order parameter has reversed from the initial $\Psi = -1$ to $\Psi > 0$ for $y \lesssim 20$~nm, while the potential is recovering towards the initial double-well with $r = -1$ (c). At $t=0.4$~ps the potential at $y=13$~nm has recovered the double-well structure (green curve in Fig.~\ref{fig:1}d) and $\Psi$ does not have sufficient kinetic energy to cross the barrier back to the negative side. 
In contrast, the potential near the surface has not developed the double-well structure yet (purple trace at $t = 0.4$~ps) and also $\Psi$ has enough energy to complete a second flip back to the $\Psi = -1$ side. 
At $t=0.7$~ps the double well starts to develop the double minimum also at the surface, eventually freezing the order parameter in the $\Psi=-1$ side at the surface. 
Finally at $t = 4$~ps, $r = -1$ everywhere and $\Psi$ freezes with two domain walls. The number of final domain walls depends on the initial strength of the excitation, $\eta$. As shown here, $\eta \approx 2$ produces two domain walls; for $\eta \approx 1$ only one domain wall forms, and for $\eta < 1$ no defects form since $r < 0$ everywhere in this case. 
Finally, the observed diffraction intensity is proportional to the CDW structure factor
\begin{equation}
	S(q_\parallel,t) = \left| \int_0^\infty \Psi(y, t) e^{-y/y_0} e^{i q_\parallel y} dy \right|^2,
\end{equation}
where $q_\parallel$ is the wavevector along $y$ and $y_0$ is the x-ray penetration depth at grazing incidence (see Supplemental Information). Fig.~\ref{fig:1}e shows a drastic decrease of $S(q_\parallel, t)$ at the nominal CDW Bragg condition, $q_\parallel=0$, but it also broadens suddenly at $0.4$~ps, demonstrated by the strong shoulders away from $q_\parallel=0$. While the peak shape recovers slightly, it remains distorted and suppressed at times $t > 4$~ps.

\begin{figure}[htb]
\centering 
\includegraphics[width=\columnwidth]{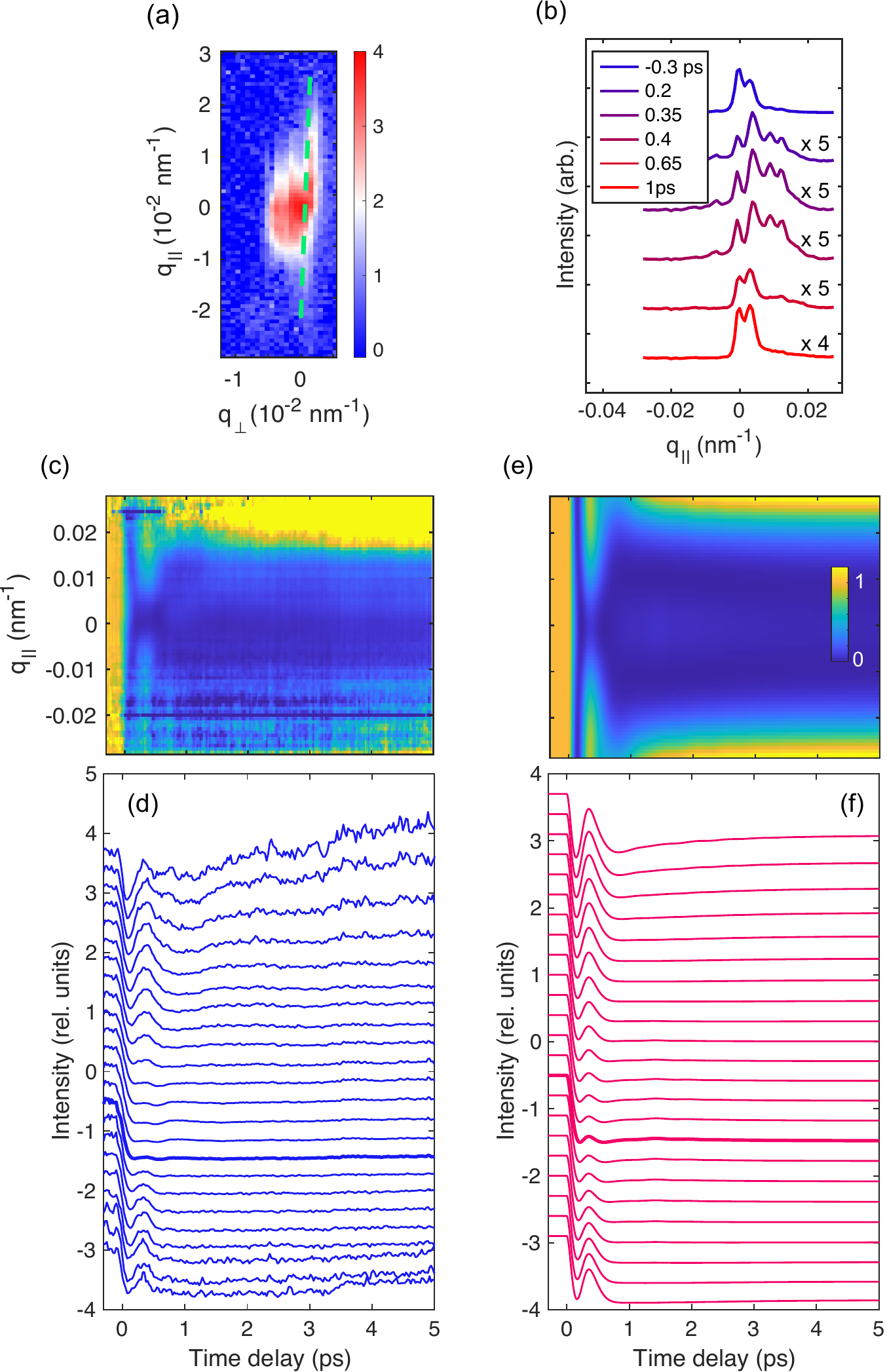} 
\caption[]{(a) Detector image (log10 intensity scale) of the $(2, 2, 1-q)$ CDW peak at room temperature taken at grazing incidence angle of $\alpha = 0.3$~deg. We define $\mathbf{q}_\parallel = (0, q_\parallel, 0)$ parallel to the sample normal ($b$-axis) and $\mathbf{q}_\perp = q_\perp(0.75,    0,    0.66)$, in the plane of the sample surface. (b) profiles of the peak for representative delays along the dashed green lines in (a) for incident fluence of $1$~mJ/cm$^2$. (c) contour plot of $\tilde{S}(q_\parallel,t)$ defined in the text at wavevectors marked by the dashed line in (a). (d) individual traces of (c) evenly-spaced between $q_\parallel = -0.013$~nm$^{-1}$ (bottom trace) and $q_\parallel = 0.02~\mathrm{nm}^{-1}$ (top trace), displaced vertically for clarity. (e) and (f) Calculated $\tilde{S}(q_\parallel,t)$ at the same wavevectors as in (c) and (d), respectively for $\eta = 2$. The traces for $q_\parallel = 0$ in (d) and (f) are indicated with a thicker line. 
\label{fig:2} } 
\end{figure}

Room temperature experiments using $9.5$~keV x-ray pulses were carried out at the X-ray Pump-Probe (XPP) station at the LCLS~\cite{chollet2015}. Grazing incidence diffraction with $0.3 < \alpha < 0.5$~deg, where $\alpha$ is the angle between the incident x-ray beam and the sample surface, was used to limit the x-ray penetration depth to $y_0 < 50$~nm. (additional details in Supplementary Information and in Ref.~\cite{trigo2019}). 
Figure \ref{fig:2}a shows a static image of the $(2,2,1-q)$ CDW sideband ($\log_{10}$ scale). This CDW  peak is mostly in-plane, the vertical direction on the image is nearly along the $b$ axis, $\mathbf{q}_\parallel=(0,q_\parallel,0)$. The horizontal detector direction is $\mathbf{q}_\perp = q_\perp \times (0.75,\;0,\; 0.66)$. 
The peak is elongated in the $b$ direction even before the pump strikes, a signature that the correlation length along the $b$-axis is shorter than in the $a-c$ plane~\cite{ru2008}. Fig.~\ref{fig:2}b shows the $q_\parallel$ dependence of the peak for representative delays at wavevectors along the widest part of the peak, indicated by the dashed line in Fig.~\ref{fig:2}a. There is a slight shift in the peak in $\mathbf{q}_\perp$, either due to a change in magnitude or direction of the wavevector~\cite{lebolloch2016effect} (Supplementary Information). The incident excitation fluence for these data was 1 mJ/cm$^2$~\cite{trigo2019}. The apparent fast oscillatory structure in Fig.~\ref{fig:2}b is likely related to preexisting domains deep beneath the surface, which do not seem affected by the pump. 
Since the total intensity is almost completely suppressed by the pump, the traces for $t > 0$ are scaled as indicated in the figure to increase visibility. Fig.~\ref{fig:2}b shows changes to the peak shape as well as intensity, particularly between $0 < t < 0.4$~ps, which seems to recover at $t > 0.65$~ps albeit with a much lower intensity (see scaling factors and Fig.~\ref{fig:2}b). To better visualize the dynamics we normalize the $\mathbf{q}_\parallel$ profiles to the average at $t < -0.1$~ps (indistinguishable from the unpumped profile). 
In Fig.~\ref{fig:2}c we show a color intensity plot of the normalized structure factor $\tilde{S}({q}_\parallel, t) = S(q_\parallel, t0)/S(q_\parallel, t<0)$ for the same wavevectors as in (b) and in Fig.~\ref{fig:2}d we plot representative intensity-vs-time traces of the same data. The normalization of $\tilde{S}(q_\parallel,t)$ removes the static modulation of the peak and brings out the time-dependent changes as can clearly be seen in (c) and (d). 
At $t\approx0$~ps the intensity is almost completely suppressed followed by a peak in $\tilde{S}(q_\parallel,t)$ at $t = 0.4$~ps for wavevectors $|q_\parallel| > 0.005$~nm$^{-1}$, and a slow increase of the intensity for these wavevectors at later times. Since $\tilde{S}$ is normalized, this indicates a sudden increase in the width of the diffraction peak at $\sim 0.4$~ps that partially relaxes back and changes slowly after $t > 0.5$~ps. This broadening of the peak is a signature of inhomogeneous dynamics in $\Psi(y,t)$ and is consistent with the schematic shown in Fig.~\ref{fig:1}e, which predicts a split and broadened peak in $S(q_\parallel,t)$ at $t \sim 0.4$~ps (shaded area in  Fig.~\ref{fig:1}e).

We use the 1D model described above as qualitative guide to understand the features observed in $\tilde{S}(q_\parallel,t)$. Fig. \ref{fig:2}e and \ref{fig:2}f show the simulated $\tilde{S}(q_\parallel,t)$ over the same wavevectors as in (c) and (d) with $y_0 = 14$~nm and $\eta = 2$ corresponding to an incident fluence of $1$~mJ/cm$^2$ in the experiment~\cite{trigo2019}  (see Supplemental Information for details). The qualitative agreement is remarkable: a peak at wavevectors $|q_\parallel| > 0.005$~nm$^{-1}$ at $t \sim 0.4$~ps, and later a slow, gradual increase of the normalized intensity at high wavevectors. A few representative snapshots of $\Psi(y,t)$ are shown in Fig.~\ref{fig:1}b and \ref{fig:1}c, with the final configuration at $t = 4$~ps containing two domain walls at $y \sim 5$~nm and $y \sim 15$~nm. Although domain walls are not topologically stable in an incommensurate CDW (they are destroyed by phase fluctuations), in $R$Te3 they seem fairly robust and exist for up to ns after the pump~\cite{moore2016,zhou2019nonequilibrium}. 
The suppression of the Bragg peak intensity in Fig.~\ref{fig:2}, a measure of the CDW long range order, is a consequence of the destructive interference between the x-rays scattered from domains with opposite sign of $\Psi$. 
This explains why the diffraction intensity is suppressed much longer~\cite{trigo2019,moore2016,zhou2019nonequilibrium} than the recovery of the local electronic order, which affects the CDW gap~\cite{zong2019}, the optical reflectivity~\cite{yusupov2008,trigo2019} and the coefficient $r(y,t)$ in the potential energy. 
We emphasize that the domain walls lie at $y\sim 5$~nm and $y\sim 15$~nm, and are likely to be present in ultrafast electron diffraction experiments~\cite{zong2019,kogar2019,zhou2019nonequilibrium} on samples thicker than the optical penetration depth $y_p \sim 20$~nm.

\begin{figure*}[htb]
\centering 
\includegraphics[width=\textwidth]{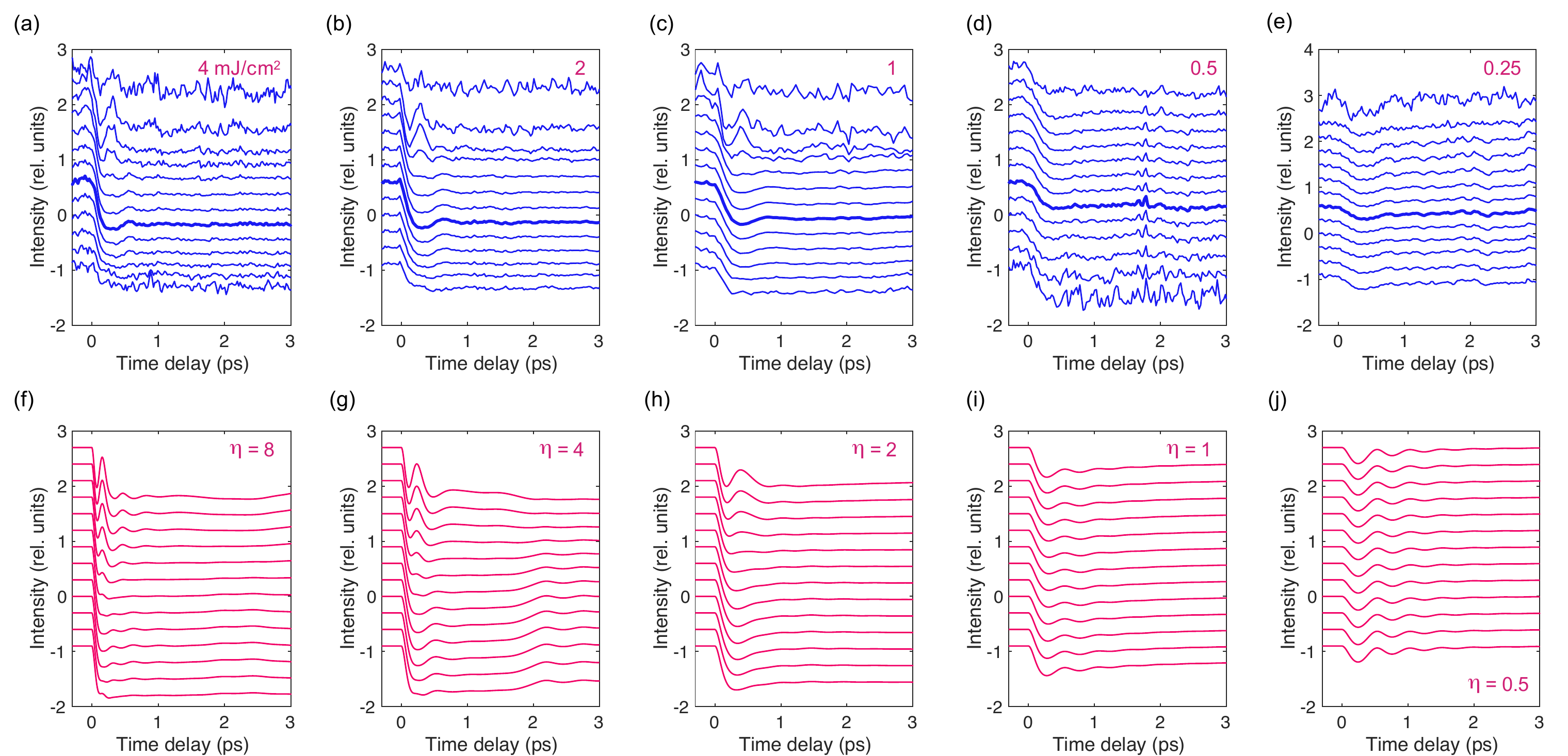} 
\caption[]{ (a-e), Dynamics of the $(1,7,q)$  peak at incident fluences of 4, 2, 1, 0.5 and 0.25 mJ/cm$^2$. (f-j), Simulation of $\tilde{S}(q_\parallel, t)$ for $\eta = 8$, 4, 2, 1 and 0.5, corresponding to experimental panels (a-e), respectively. All the experimental traces were taken with an x-ray incidence angle of $0.5$~deg. The wavevector for the nominal Bragg condition is indicated with a thick line and the traces are separated by $1.49 \times 10^{-3}$~nm$^{-1}$ and have been displaced vertically for clarity. The spurious spike at $1.8$~ps in (d) is due to a glitch in the x-ray source.
\label{fig:3} } 
\end{figure*}

We now turn to the fluence dependence of $\tilde{S}(q_\parallel,t)$, summarized in Fig. \ref{fig:3} for the $(1,7,q)$ CDW peak measured at an incidence angle of $\alpha = 0.5$~deg (a-e), and the corresponding simulation (f-j). 
The traces correspond to wavevectors separated by $1.49 \times 10^{-3}$ nm$^{-1}$ along the vertical direction on the detector and are displaced vertically for clarity. 
These wavevectors have a small projection in the a-c plane since $(1,7,q)$ has a larger out-of-plane component.
We find good qualitative agreement between the model and the experimental data. In particular, the peaks at $t < 0.5$~ps for the top traces away from the nominal Bragg condition are well reproduced over all the fluences $> 0.5$~mJ/cm$^2$ (a-c) and (f - h). 
Importantly, this peak does not appear for fluences $\leq 0.5$~mJ/cm$^2$ (d and e) which agrees with the simulation for $\eta \leq 1$ (i and j). 
For $\eta = 1$, $r(0,0) = 0$, a regime associated with dynamical slowing down~\cite{zong2019dynamical}, thus $\Psi$ has a small kinetic energy and flips only once, producing a single domain wall.
The overall intensity is suppressed by the domain wall, but there are no oscillations. 
Finally, no domain wall are produced for lower excitation $\eta < 1$ (e and j). In this case, the intensity recovers within a ps after a short, nearly harmonic transient due to the coherent dynamics of the amplitude mode of the CDW~\cite{yusupov2008,schmitt2008,chang2012,rettig2014,leuenberger2015,trigo2019}.

\begin{figure*}[htb]
\centering 
\includegraphics[width=\textwidth]{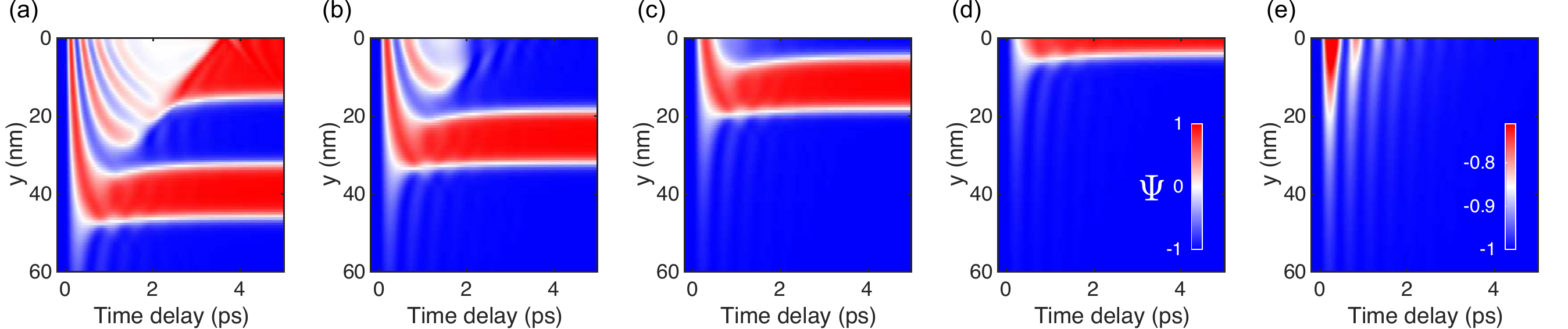} 
\caption[]{(a-e) contour plots of $\Psi(y,t)$ for $\eta = 8$, 4, 2, 1 and 0.5 (a-e), respectively. The color scale for (a-d) is shown in (d).
\label{fig:4} } 
\end{figure*}

The contour plots in Figure \ref{fig:4} show the calculated dynamics of $\Psi(y,t)$ for excitations of $\eta = 8$, 4, 2, 1 and 0.5 (a-e) matching those of Fig.~\ref{fig:3}. Blue (red) corresponds to $\Psi < 0$ ($\Psi > 0$). For $\eta > 1$ the dynamics produces one (d), two (b-c) or three (a) domain walls, whose locations along the depth (vertical axis) depend on $\eta$. At $\eta = 0.5$ not only does $\Psi(y,t)$ not flip to $\Psi > 0$, but it behaves as a nearly-harmonic oscillator whose frequency is slightly chirped with longer period near the surface, which recovers to the equilibrium $\Psi = -1$ in less than 4 ps (Fig.~\ref{fig:4}e). In the limit of small $\eta$, $\Psi(y,t)$ is harmonic around the initial potential minimum and the dynamics of the CDW peaks reflect the coherent dynamics of the amplitude mode of the CDW~\cite{trigo2019}. 
Finally, for $\eta = 8$ and $\eta= 4$ (Fig.~\ref{fig:4}a and \ref{fig:4}b),  $r(0,t) > 0$ for $0<t<2$ ps, and the potential at the surface, $y=0$, is quadratic for sufficiently long time that $\Psi(y=0,t)$ can perform several harmonic oscillations around the quadratic potential with $r(0,t) > 0$  (with minimum at $\Psi = 0$) as can be seen in Fig.~\ref{fig:4}a and \ref{fig:4}b near the surface ($y=0$) and for $t < 2$~ps. This motion results in low-frequency oscillations in the diffraction data at $0 < t < 2$~ps, most clearly seen at $\alpha = 0.4$~deg (supplementary information).

Using ultrafast x-ray diffraction with an XFEL, we showed how photoexcitation generates non-trivial configurations of the order parameter in a charge ordered system in the form of domain walls propagating perpendicular to the sample surface. These domain walls break the CDW long range order and suppress the diffraction intensity of the CDW for times much longer than the recovery of the local electronic gap.
These features are produced and measured stroboscopically over multiple repetitions of pump-probe pulses and must therefore be generated in a deterministic manner. This ability to produce defects on demand and to image their dynamics will provide a more complete picture of the competition between the nearly degenerate $c$- and $a$-axis orders in $R$Te$_3$, which can be lifted by photoexcitation~\cite{kogar2019}, and may pave the way towards better understanding of other coupled broken symmetries in the $R$Te$_3$ system~\cite{hamlin2009} and other systems with competing orders~\cite{chang2012,fradkin2015theory}.


Preliminary x-ray characterization was performed at BL7-2 at the Stanford Synchrotron Radiation Lightsource (SSRL). MK, TH, MT, DL, PSK, ZXS, PG-G, IRF and DAR were supported by the U.S. Department of Energy, Office of Science, Office of Basic Energy Sciences through the Division of Materials Sciences and Engineering under Contract No. DE-AC02-76SF00515. Use of the LCLS and SSRL is supported by the U.S. Department of Energy, Office of Science, Office of Basic Energy Sciences under Contract No. DE-AC02-76SF00515. JNC was supported by the Volkswagen Foundation. Additional X-ray measurements were performed at BL3 of SACLA with the approval of the Japan Synchrotron Radiation Research Institute (JASRI) (Proposal No. 2016A8008).

%


\end{document}